\documentclass[12pt]{article}
\input{epsf}

\newcommand{\be}{\begin{equation}}
\newcommand{\ee}{\end{equation}}
\newcommand{\ba}{\begin{eqnarray}}
\newcommand{\ea}{\end{eqnarray}}

\newcommand{\beq}{\begin{eqnarray}}
\newcommand{\eeq}{\end{eqnarray}}
\newcommand{\ket}{\rangle}
\newcommand{\bra}{\langle}

\newcommand{\del}{\partial}

\newcommand{\half}{\frac{1}{2}}

\newcommand{\Bmass}{M} 
\newcommand{\ResMass}{M_{R}} 
\newcommand{\protonmass}{M_{p}} 

\newcommand{\Mmass}{m} 
\newcommand{\Rscal}{\mu} 
\newcommand{\fpi}{f} 

\newcommand{\delmu}{\partial_\mu}

\begin{document}
\begin{center}
{\Large{\bf Magnetic moments of the $\Lambda$(1405) and $\Lambda$(1670)
  resonances
  }}

\vspace{0.3cm}

\end{center}

\vspace{1cm}

\begin{center}
{\large{D. Jido$^{1)}$, A. Hosaka$^{2)}$, J.C. Nacher$^{3)}$,
E. Oset$^{3)}$ and A. Ramos$^{1)}$}}
\end{center}

\begin{center}
{\small{$^{1)}$ \it 
     Departament d'Estructura i Constituents de la Mat\`{e}ria,\\ 
     Universitat de Barcelona, 
     Diagonal 647, 08028 Barcelona, Spain
}}

{\small{$^{2)}$ \it 
     Research Center for Nuclear Physics (RCNP), Osaka University,\\ 
     Ibaraki, Osaka 567-0047, Japan
}}

{\small{$^{3)}$ \it 
     Departamento de F\'{\i}sica Te\'orica and IFIC, \\
     Centro Mixto Universidad de Valencia-CSIC, \\
     Ap. Correos 22085, E-46071 Valencia, Spain }}

\end{center}

\vspace{1cm}

\begin{abstract}
    By using techniques of unitarized chiral perturbation theory, 
    where the $\Lambda(1405)$ and $\Lambda(1670)$ resonances are 
    dynamically generated, we evaluate the magnetic moments of these 
    resonances and their transition magnetic moment. The results
    obtained here differ appreciably from those obtained with existing
    quark models. The width for the $\Lambda(1670) \rightarrow
    \Lambda(1405) \gamma$ transition is also evaluated leading to a
    branching ratio of the order of $2 \times 10^{-6}$. 
\end{abstract}

\section{Introduction}
The evaluation of static properties of baryonic resonances, like the
magnetic moment, is a standard exercise when one has a wave function
for the states. This is the case of the quark models where a thorough
investigation of magnetic moments and other static
properties \cite{isgurkarl}, such as masses and 
couplings to the $\pi N$ system \cite{pedro}, has been done.

The introduction of unitary chiral techniques has allowed one to show that
the octet of the lowest energy $J^P = 1/2^-$ baryonic resonances can be
generated dynamically from the lowest order chiral Lagrangian and by 
the
use of natural size cut-offs or regularizing scales to make the
divergent loop integrals finite.  
These findings allow one to classify those states as
quasibound meson-baryon states, or equivalently, ordinary multiple
scattering resonances in coupled channels.  The $\Lambda(1405)$ 
was one of the first resonances to receive attention from the chiral
unitary perspective \cite{Kai95, angels, joseulf}.  The $N^*(1535)$
was also generated within chiral unitary schemes in \cite{Kai97,
Nacher:2000vg,Nieves:2001wt} and has been recently revised in
\cite{Inoue:2001ip} with the inclusion of $\pi\pi N$ channels.   
Recently the $\Lambda(1670)$ and $\Sigma(1620)$ \cite{cornelius} and
the $\Xi(1620)$ states \cite{angelsnow} have also been generated
within the same scheme, thus completing the octet of dynamically
generated states.

In the chiral unitary method, one computes scattering matrices 
in all meson and baryon channels and poles for resonances are searched 
in the second Riemann sheet.  The poles provide the mass and the width
of the resonance states and, in addition, 
the residues at the poles provide the product
of the couplings of the resonance to the initial and final states of
the considered transition scattering matrix element.  
In this method, it is not straightforward to
evaluate other properties of the resonance, like magnetic moments,
since, unlike ordinary quantum mechanical problems, 
in the present approach we do not have wave functions 
and operators manifestly.  
Therefore, we need to explore an alternative  method  
to compute resonance magnetic moments from scattering matrices. 
This is the subject of the present work.  
We compute the  magnetic moments of the $\Lambda(1405)$ and
$\Lambda(1670)$ resonances, as well as the transition magnetic moment
from the $\Lambda(1670)$ to the $\Lambda(1405)$, which allows us to
determine the partial decay width for the decay $\Lambda(1670)
\rightarrow \Lambda(1405) \gamma$.  We also compare the results
obtained here with those of ordinary quark models showing that there
are appreciable differences between them.  This offers an evidence
that the nature of these states as dynamically generated from multiple
scattering of coupled channels of mesons and baryons differs from 
an ordinary quark model description.

The paper is organized as follows.  
In section 2, we briefly describe the model that we use and show 
in detail the method to compute scattering matrices.  
In section 3, we compare the scattering matrices with a resonance 
dominant form and extract the magnetic moments.  
In section 4, we present our numerical results, which are compared with 
quark model results in section 5.  
The final section summarizes our findings.

\section{Evaluation of the magnetic moment}

The procedure to evaluate the magnetic moment of the resonances
proceeds in an analogous way to that for the $N^*N^*\pi$
coupling in \cite{Nacher:2000vg}.  We evaluate the $T$-matrix for the
process $M B \rightarrow M^\prime B^\prime \gamma$
using the chiral Lagrangian
for the coupling of the mesons and baryons and for the photon to the
mesons and baryons.  We sum the Feynman diagrams which generate 
the resonance both on the left and on the right of the photon coupling.
Isolation of resonance poles from these diagrams 
then allows us to evaluate the resonance magnetic moment.

\begin{figure}
    \epsfxsize=14cm
    \epsfbox{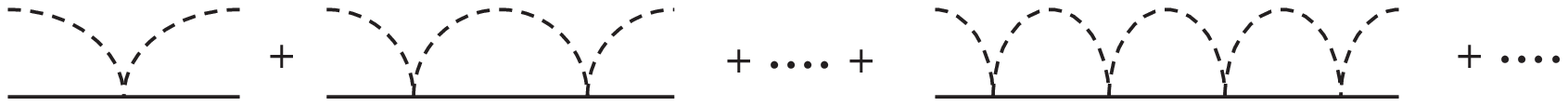}
    \caption{Diagrammatic representation of the Bethe-Salpeter equation
    in eqs.\ (\ref{BSeq}) and (\ref{inv}). Dashed and solid lines
    denote the meson and the baryon, respectively. \label{fig:T-amp}}
 \end{figure}

The $\Lambda(1405)$ resonance is generated in \cite{angels} by means
of the Bethe-Salpeter equation with a cut-off to regularize the loop 
integrals. 
The Bethe-Salpeter equation is given by
\begin{equation}
    T = V + VGT \ , \label{BSeq}
\end{equation}
where in the present method 
the term $VGT$ is given as a matrix product of the potential $V$, the
meson-baryon propagator $G$ and 
the $T$-matrix $T$.  
The diagonal matrix $G$ contains the loop integral of a meson and  baryon
propagators.  
In general, the product $VGT$ involves an integral over off-shell 
momenta.  
In the present approach that integral is greatly simplified  
reducing the problem to a  matrix product due to the on-shell
factorizations of $V$ and $T$.  
The on-shell factorization in \cite{angels} was
done by incorporating the off-shell part of the loops into
renormalization of couplings of the lowest order Lagrangian, in analogy
to what was done in the meson-meson interaction in
\cite{Oller:1997ti}.  An explicit demonstration of the cancellation of
these terms with tadpole corrections can also be seen in \cite{daniel}
for the $p$-wave meson-meson interaction in the $\rho$ channel.  The
on-shell factorization allows one to solve eq.(\ref{BSeq}) to give 
\begin{equation}
    T = [1-VG]^{-1} V\ , \label{inv}
\end{equation}
in a simple matrix inversion.  
This has also been derived using the unitarization with the $N/D$
method and dispersion relations in \cite{joseulf}.  In this latter
paper \cite{joseulf} 
the regularization of the loops is done by means of dimensional
regularization with subtraction constants in the $G$ function. 
The same method was used in \cite{cornelius} to obtain the $\Lambda(1405)$
and $\Lambda(1670)$ resonances, which is the one we follow here.
The Feynman diagrams summed by eqs.\ (\ref{BSeq}) and (\ref{inv}) are
given in Fig.\ref{fig:T-amp}.  

The $s$-wave meson-baryon interaction potential $V$ is
derived from the second order
terms in the meson field of the 
chiral Lagrangian \cite{Be95,Pi95}:
\begin{equation}
   V_{ij} = - C_{ij} {1 \over 4 \fpi^2} (2 \sqrt{s} -\Bmass_i -
   \Bmass_j) \left( { \Bmass_i + E \over 2 \Bmass_i} \right)^{1/2}
   \left( { \Bmass_j + E^\prime \over 2 \Bmass_j} \right)^{1/2}  \ ,
   \label{Vpot} 
\end{equation}
where the coefficients $C_{ij} (= C_{ji})$ are given in \cite{angels}
and the meson decay constant $f$ is taken as an average value
$f=1.123 f_\pi$. 
The $G$ function for each meson-baryon channel is given by
\begin{eqnarray}
    G_l(\sqrt{s}) 
    &=& i 2 \Bmass_l \int {d^4 q \over (2\pi)^4 } { 1 \over
    (P-q)^2 - \Bmass_l^2 + i \epsilon}{1 \over q^2 - \Mmass_l^2 +
    i\epsilon} \nonumber \\
    &=& {2 \Bmass_l \over 16 \pi^2} \left\{ a(\Rscal) + \ln {\Bmass^2_l
    \over \Rscal^2} + {\Mmass_l^2 - \Bmass_l^2 + s \over 2 s} \ln
    {\Mmass^2_l \over \Bmass^2_l} \right. \\ \nonumber 
    && + {\bar q_l \over \sqrt{s} } \left[ \ln(s - (\Bmass^2_l -
    \Mmass^2_l) + 2 \bar q_l \sqrt{s}) + \ln(s + (\Bmass^2_l -
    \Mmass^2_l) + 2 \bar q_l \sqrt{s}) \right. \nonumber \\
    && \left. \left.  - \ln(- s + (\Bmass^2_l - \Mmass^2_l) + 2 \bar
    q_l \sqrt{s}) - \ln(- s - (\Bmass^2_l - \Mmass^2_l) + 2 \bar q_l
    \sqrt{s}) \right] \right\} \ , \nonumber
\end{eqnarray}
where $\Mmass$ and $\Bmass$ are taken to be the observed meson and baryon
masses, respectively, and $\mu$ is a regularization scale which is
chosen to be 630 MeV as in \cite{cornelius}.  
The subtraction constants $a_l$ are of the order of $-2$, 
which is a natural size as shown in \cite{joseulf}.  
The values chosen in \cite{cornelius},
which reproduce the results of \cite{angels} calculated with just one
cut-off, are
\begin{eqnarray}
   a_{\bar K N} = -1.84, \hspace{0.5cm} a_{\pi \Sigma} = - 2.00,
   \hspace{0.5cm} a_{\pi \Lambda} = -1.83 \nonumber \\
   a_{\eta \Lambda} = -2.25, \hspace{0.5cm} a_{\eta \Sigma} = -2.38,
   \hspace{0.5cm} a_{K \Xi} = -2.67
\end{eqnarray}

\begin{figure}
    \epsfxsize=14cm
    \epsfbox{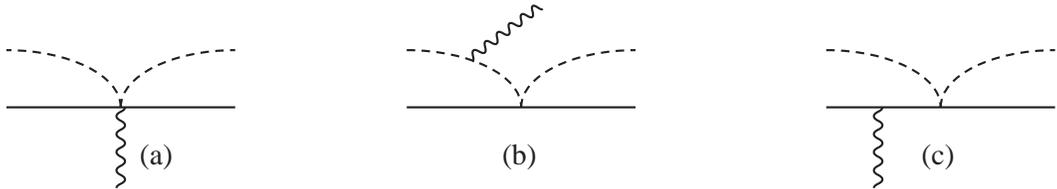}
    \caption{The elementary couplings of the photon to the components of the
    meson-baryon amplitude. The wavy line denotes the photon.\label{fig:lowest}}
 \end{figure}

\begin{figure}
    \epsfxsize=14cm
    \epsfbox{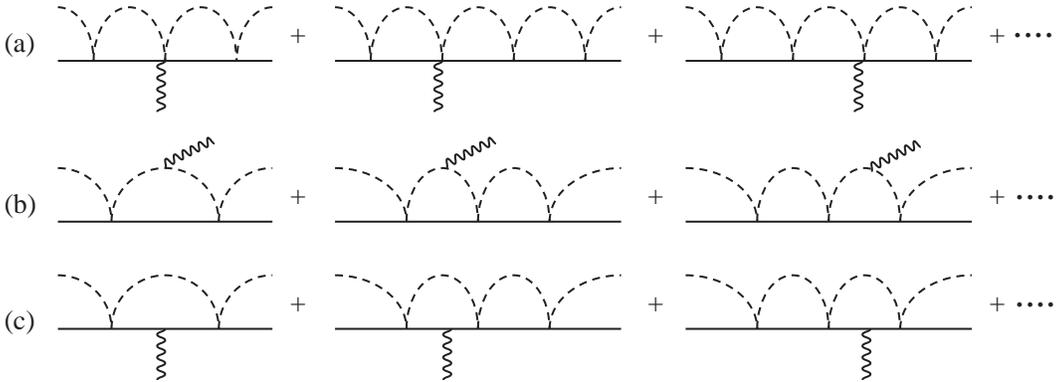}
    \caption{Diagrams for the coupling of the photon to the resonance
    dynamically generated in meson-baryon scattering. \label{fig:res}}
 \end{figure}

The elementary couplings of the photon to the components of the
meson-baryon amplitude at lowest order of the chiral expansion 
are shown in
fig.\ref{fig:lowest}.  Now if we want to generate the resonance on the
left and right sides of the photon coupling we must consider the diagrams
shown in fig.\ref{fig:res}.  The diagrams of row b) in
fig.\ref{fig:res} vanish, given the $s$-wave nature of the meson-baryon
vertices and the $\vec \epsilon \cdot \vec q_{L}$ coupling of the
photon to the mesons, which makes the integral over the loop variable
$q_{L}$ vanish.  The remaining couplings are those of the photon to the
baryons and the analogous ones with two extra meson lines.  
The spin
dependent part of these couplings needed for the evaluation of 
magnetic moments
is given by \cite{Meissner:1997hn}
\begin{equation}
   {\cal L} = - {i \over 4 \protonmass} b^F_6 \langle \bar B
[S^\mu,S^\nu ]
   [F^+_{\mu\nu},B] \rangle - {i \over 4 \protonmass} b^D_6 \langle
\bar B
   [S^\mu,S^\nu ] \{F^+_{\mu\nu},B\} \rangle \label{mag}
\end{equation}
with 
\begin{eqnarray}
   F^+_{\mu\nu} &=& - e (u^\dagger Q F_{\mu\nu} u + u Q F_{\mu\nu}
   u^\dagger) \label{photon} \\
   F_{\mu\nu} &=& \delmu A_\nu - \partial_\nu A_\mu \ ,
\end{eqnarray}
where $\protonmass$ is the mass of proton, $A_\mu$ is the electromagnetic 
field, and $b^F_6$ and $b^D_6$ are parameters to be fitted so as to
reproduce the magnetic moments of the ground state baryons.
In eq.(\ref{mag}), $\langle \cdots \rangle$ means the trace over 
flavor indices,  $B$ is the SU(3)
matrix for the baryon field \cite{Be95,Pi95}, 
and $S^\mu$ are spin matrices as explained below.  
In eq.(\ref{photon}) $Q$ is the charge matrix for the $u, d, s$ quarks: 
$Q=
{1 \over 3} {\rm diag} (2,-1,-1)$ and $u^2 = U = \exp(i \sqrt{2} \Phi/
\fpi)$ where $\Phi$ is the SU(3) matrix of the pseudoscalar meson
field \cite{Be95,Pi95,gl}.   
In the baryon rest frame the operator $S^\mu$ becomes $\vec\sigma /2$
and then,
\begin{equation}
   [S^\mu, S^\nu] F_{\mu\nu} \rightarrow - (\vec \sigma \times \vec
   q\, ) \cdot \vec \epsilon 
\end{equation}
in the Coulomb gauge ($\epsilon^0=0$, $\vec \epsilon \cdot \vec q=0$)
and for an outgoing photon.  Thus the vertex from the Lagrangian of
eq.(\ref{mag}) can be written as
\begin{eqnarray}
   {\cal L} &\rightarrow&  e{\vec\sigma \times \vec q \over 2
   \protonmass}\cdot \vec \epsilon  \left( - {i \over 2} b^F_6 \left\langle
   \bar B [(u^\dagger Qu+uQu^\dagger),B] \right\rangle \right. \\
   &&\ \  \left. - {i \over 2} b^D_6 \left\langle \bar B \{(u^\dagger
   Qu+uQu^\dagger),B\} \right\rangle \right) \ .
\end{eqnarray}
By expanding $u$ in terms of the meson field we obtain the expressions
for both the $\gamma BB^\prime$ and $\gamma BB^\prime MM^\prime$
vertices.  
By taking $u=1$ we obtain the magnetic moments of the ground state octet 
baryons, 
\begin{equation}
   \mu_i = d_i b_6^D + f_i b_6^F \label{magmon}\ ,
\end{equation}
where the coefficients $d_i$ and $f_i$ are given in table
\ref{tab:magmon}.  
One immediately realizes that by 
setting $b_6^D=0$ and $b_6^F=1$ one obtains
the ordinary magnetic moments of the baryons without anomalous 
contributions.  
Fitting the values of
eq.(\ref{magmon}) to the observed magnetic moments of the baryons one
obtains
\begin{equation}
   b_6^D = 2.40, \hspace{1cm} b_6^F = 1.82
\end{equation}
very similar to those given in \cite{Meissner:1997hn}, $b_6^D = 2.39,
b_6^F = 1.77$.

\begin{table}
  \begin{center}
   \begin{tabular}{l|ccccccccc}
        & $p$ & $n$ & $\Sigma^+$ & $\Sigma^-$ & $\Sigma^0$ & $\Lambda$ &
         $(\Lambda \Sigma^0)$ & $\Xi^-$ & $\Xi^0$ \\
     \hline
     $d_i$ & $1 \over 3$ & $-{ 2 \over 3}$ & $1 \over 3$ & $1 \over 3$
        & $1 \over 3$ & $-{1 \over 3}$ & $1 \over \sqrt 3$ & $1 \over
        3$ & $-{2 \over 3}$ \\
     $f_i$ & $1$ & $0$ & $1$ & $-1$ & $0$ & $0$ & $0$ & $-1$ & $0$
   \end{tabular}
   \caption{$d_i$ and $f_i$ coefficient of
   eq.(\ref{magmon}). \label{tab:magmon}}
  \end{center}
\end{table}

Similarly, by expanding $u$ up to two meson fields we obtain the
vertices of diagram a) of fig.\ref{fig:lowest} with the result
\begin{equation}
   -i t_{ij}^{a)} = 
    {e \over 2 \protonmass} (\vec \sigma \times \vec q \,)
    \cdot \vec \epsilon {1 \over 2 \fpi^2} [ X_{ij} b_6^D + Y_{ij} b_6^F ]
    \ ,  \label{xyterms}
\end{equation}
where the coefficients $X_{ij}$ and $Y_{ij}$ are given in tables
\ref{tab:xcoff} and \ref{tab:ycoff}.

\begin{table}
  \begin{center}
  \begin{tabular}{c|rrrrrrrrrr} 
        & $K^-p$ & $\bar K^0 n$ & $\pi^0 \Lambda$ & $\pi^0 \Sigma^0$ &
        $\eta \Lambda$ & $\eta \Sigma^0$ &
         $\pi^+ \Sigma^-$ & $\pi^- \Sigma^+$ & $K^+\Xi^-$ & $K^0\Xi^0$ \\
     \hline
     $K^-p$ & $0$ & $-{ 1 \over 2}$ & $ - {1 \over 4 \sqrt 3}$ & $1
\over 4$
        & $-{1 \over 4}$ & ${3 \over 4 \sqrt 3}$ & $0$ & $1$ & $0$ & $0$ \\
     $\bar K^0 n$ & & $0$ & $0$ & $0$ & $0$ & $0$ & $- {1\over 2}$ &
        $0$ & $0$ & $0$ \\
     $\pi^0 \Lambda$ & & & $0$ & $0$ & $0$ & $0$ & $ {1\over \sqrt 3}$ &
        $1 \over \sqrt 3$ & $-{1 \over 4 \sqrt 3}$ & $0$ \\
     $\pi^0 \Sigma^0$ & & & & $0$ & $0$ & $0$ & $0$ & $0$ & 
        $1 \over 4$ & $0$ \\
     $\eta \Lambda$ & & & & & $0$ & $0$ & $0$ & $0$ & 
        $-{1 \over 4}$ & $0$ \\
     $\eta \Sigma^0$ & & & & & & $0$ & $0$ & $0$ & $3 \over 4 \sqrt 3$
        & $0$ \\
     $\pi^+ \Sigma^-$ & & & & & & & $0$ & $0$ & $1$ & $0$ \\
     $\pi^- \Sigma^+$ & & & & & & & & $0$ & $0$ & $-{1 \over 2}$ \\
     $K^+ \Xi^-$ & & & & & & & & & $0$ & $-{1 \over 2}$ \\
     $K^0 \Xi^0$ & & & & & & & & & & $0$ \\
  \end{tabular}
  \caption{$X_{ij}$ coefficient of eq.(\ref{xyterms}). $X_{ij}=X_{ji}$
  \label{tab:xcoff}}
  \end{center}
\end{table}

\begin{table}
  \begin{center}
  \begin{tabular}{c|rrrrrrrrrr} 
        & $K^-p$ & $\bar K^0 n$ & $\pi^0 \Lambda$ & $\pi^0 \Sigma^0$ &
        $\eta \Lambda$ & $\eta \Sigma^0$ &
         $\pi^+ \Sigma^-$ & $\pi^- \Sigma^+$ & $K^+\Xi^-$ & $K^0\Xi^0$ \\
     \hline
     $K^-p$ & $-2$ & $-{ 1 \over 2}$ & $ - {3 \over 4 \sqrt 3}$ & 
         $-{1\over 4}$ & $-{3 \over 4}$ & $-{3 \over 4 \sqrt 3}$ & 
         $0$ & $-1$ & $0$ & $0$ \\
     $\bar K^0 n$ & & $0$ & $0$ & $0$ & $0$ & $0$ & ${1\over 2}$ &
        $0$ & $0$ & $0$ \\
     $\pi^0 \Lambda$ & & & $0$ & $0$ & $0$ & $0$ & $0$ &
        $0$ & ${3 \over 4 \sqrt 3}$ & $0$ \\
     $\pi^0 \Sigma^0$ & & & & $0$ & $0$ & $0$ & $1$ & $-1$ & 
        $1 \over 4$ & $0$ \\
     $\eta \Lambda$ & & & & & $0$ & $0$ & $0$ & $0$ & 
        ${3 \over 4}$ & $0$ \\
     $\eta \Sigma^0$ & & & & & & $0$ & $0$ & $0$ & $3 \over 4 \sqrt 3$
        & $0$ \\
     $\pi^+ \Sigma^-$ & & & & & & & $2$ & $0$ & $1$ & $0$ \\
     $\pi^- \Sigma^+$ & & & & & & & & $-2$ & $0$ & $-{1 \over 2}$ \\
     $K^+ \Xi^-$ & & & & & & & & & $2$ & ${1 \over 2}$ \\
     $K^0 \Xi^0$ & & & & & & & & & & $0$ \\
  \end{tabular}
  \caption{$Y_{ij}$ coefficient of eq.(\ref{xyterms}). $Y_{ij}=Y_{ji}$
  \label{tab:ycoff}}
  \end{center}
\end{table}

The evaluation of the amplitudes corresponding to the diagrams of
fig.\ref{fig:res} (the magnetic part) is straightforward.  We obtain,
\begin{equation}
    -i t_{ij}^{\gamma} = -i\tilde{t}_{ij} {e \over 2 \protonmass} 
    (\vec\sigma \times \vec q\, )\cdot \vec \epsilon \label{gamamp}
\end{equation}
and
\begin{equation}
   -i \tilde t_{ij} = \left( \sum_{lm} t_{i l} G_l A_{lm} G_m 
   t_{mj} + \sum_l t_{il} \tilde G_l t_{lj} \mu_{B_l}
  \right) \, . \label{ttilde}
\end{equation}
In this equation 
$t_{ij}$ is the scattering amplitude from the channel $i$ to $j$,
\begin{equation}
    A_{lm}={1 \over 2\fpi^{2}} [X_{lm}b^{D}_{6} + Y_{lm}b_{6}^{F}] \ ,
\end{equation}
and
\begin{equation}
    \tilde G_{l}(p) = i \int {d^{4}k \over (2 \pi)^{4}} D(k) G(p-k)
     G(p-k) 
\end{equation}
with $D$ and $G$ the meson and baryon propagators.  
Here, by keeping up 
to linear terms in $q$, we have neglected the small momentum of the 
photon in the second baryon propagator. 
Therefore, we can write
\begin{equation}
    \tilde G_{l}(\sqrt{s}) = -{\del \over \del \sqrt s} G_{l} \ .
\end{equation}
This approximation allows us to obtain an analytic expression for
$\tilde{G}_l(\sqrt{s})$.
In eq.(\ref{ttilde}), we omit to write contributions from the
$\Lambda$-$\Sigma^0$ transition magnetic moment. The contributions are
negligible since the $\Lambda$-$\Sigma^0$ transition changes the
isospin, therefore either the left or right resonances must have isospin
1, which is not the present case.

\section{Comparison to the resonance description}

\begin{figure}
    \epsfxsize=14cm
    \epsfbox{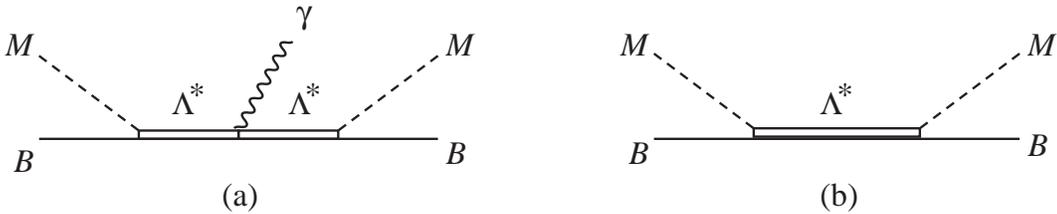}
    \caption{a) Diagrammatic representation of the photon coupling to
    an explicit resonance. b) Diagrammatic representation of
    meson-baryon scattering through the explicit resonance.
    \label{fig:resonance}}
 \end{figure}

In order to extract a resonance magnetic moment from the scattering 
amplitude, (\ref{gamamp}) or (\ref{ttilde}), we assume that 
resonances are dynamically generated on the left and right of the 
photon coupling.  
%
%
First we parameterize the meson-baryon scattering amplitude $t_{ij}$ 
as shown in fig.\ref{fig:resonance}(b) 
by the resonance dominant 
Breit-Wigner form: 
\begin{equation}
    -i t_{ij} = -i g_{i}  {i \over \sqrt s - \ResMass + i
     \Gamma/2} (-i g_j^{*})\ .
\end{equation}
Here we have introduced the resonance mass $M_{R}$, the total decay 
width $\Gamma$ and the decay constant to the channel $i$, $g_{i}$.   
Then the photon coupling amplitude $t_{ij}^{\gamma}$ is parameterized 
as shown in fig.\ref{fig:resonance}(a) by the expression:
\begin{equation}
    -i t_{ij}^{\gamma} = - i g_{i} {i \over \sqrt{s} - \ResMass + 
    i\Gamma/2 }{e  \mu_{\Lambda^{*}} \over 2 \protonmass}(\vec \sigma 
    \times \vec q \, )\cdot \vec \epsilon { i \over \sqrt s - \ResMass + 
    i \Gamma/2} (-i) g_{j}^* \label{magamp} \ .
\end{equation}
Dividing $-i t_{ij}^{\gamma}$ by $t_{ij}$ and by ${e \over 2 
\protonmass}(\vec \sigma \times \vec q\,)\cdot \vec \epsilon$ we cancel the 
coupling constants and one propagator. Thus by evaluating this ratio 
at the $\Lambda^{*}$ pole, where the amplitudes are dominated by the 
resonance, and recalling eq.(\ref{gamamp}), we have 
\begin{equation}
    \mu_{\Lambda^{*}} = \lim_{z\rightarrow z_{R}} (z-z_R) 
     {-i \tilde t_{ij}(z) 
    \over t_{ij}(z) } = \left. {\rm Res}{-i \tilde t_{ij}(z) \over
    t_{ij}(z)} \right|_{z=z_{R}} \label{compratio}\ . 
\end{equation}
%
%
where $z_R$ denotes the position of the pole in the second Riemann sheet,
$z_R \equiv M_R + i \Gamma/2 $.
In fact, there exist two poles around the region of the $\Lambda(1405)$
\cite{joseulf}, located  at $z_R
= 1426 + 16i$ and $1390+66i$ MeV. The former pole largely couples to
the $\bar K N$ state, whereas the latter one couples predominantly to
the $\pi \Sigma$ state. 
Both poles may contribute to the resonance $\Lambda
(1405)$. We evaluate the magnetic moment at both poles. For the
$\Lambda (1670)$ the pole position is $z_R = 1680 + 20i$ MeV. 

Similarly, we can also evaluate the transition amplitude between the 
$\Lambda(1670)$ and $\Lambda(1405)$ resonances. This is 
accomplished by
putting different energies, $\sqrt{s_1}$ and $\sqrt{s_2}$, on the
transition amplitudes $t_{ij}$ appearing on the left
and right of the photon coupling in eq.\ (\ref{ttilde}).
Then by  taking $\sqrt {s_{1}} \equiv z_{1R}$ for the first 
resonance ($\Lambda(1670)$) and $\sqrt{s_{2}}\equiv z_{2R}$ 
for the second resonance ($\Lambda(1405)$), 
we would find 
\begin{equation}
    \mu_{\Lambda(1670)\rightarrow \Lambda(1405)} = 
    \lim_{ \stackrel{\scriptstyle z_{1}\rightarrow z_{1R}}
	{z_{2}\rightarrow z_{2R}}} 
    {-i\tilde t_{ij}(z_{1}, z_{2}) g_{i}(1670) g_j^*(1405) \over
    t_{ii}(z_{1}) t_{jj} (z_{2})} \, . \label{tranratio}
\end{equation}

\begin{figure}
   \begin{center}
    \epsfxsize=10cm
    \epsfbox{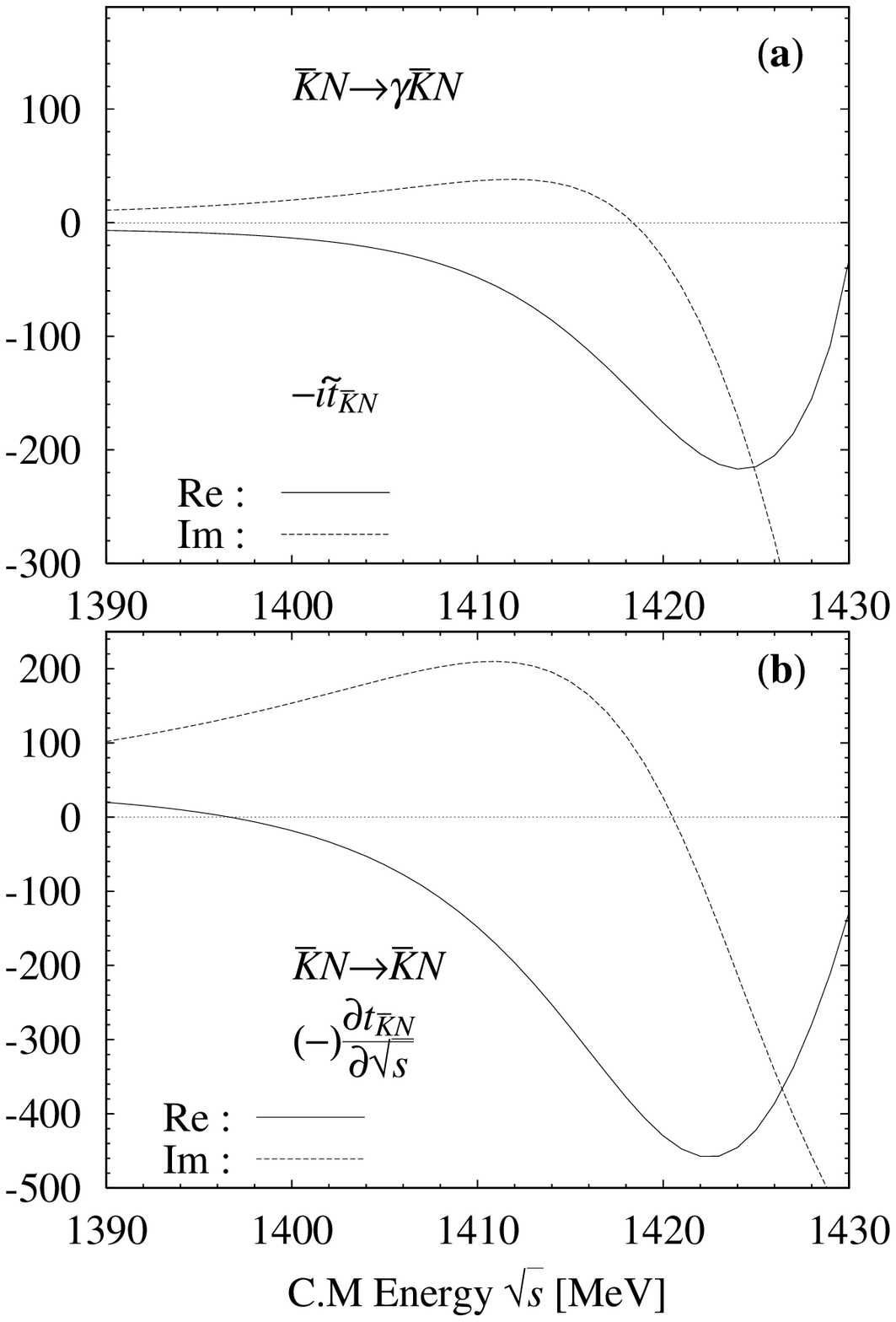}
   \end{center}
   \caption{ Real and imaginary parts of a) the numerator $-i \tilde
    t_{\bar K N}$ in eq.(\ref{realratio}) and b) the denominator $-\del
    t_{\bar K N} / \del \sqrt{s}$ in eq.(\ref{realratio}) around the
    $\Lambda(1405)$ resonance region in units of
    $m_\pi^{-2}$. \label{fig:plot1}} 
 \end{figure}
\begin{figure}
   \begin{center}
    \epsfxsize=10cm
    \epsfbox{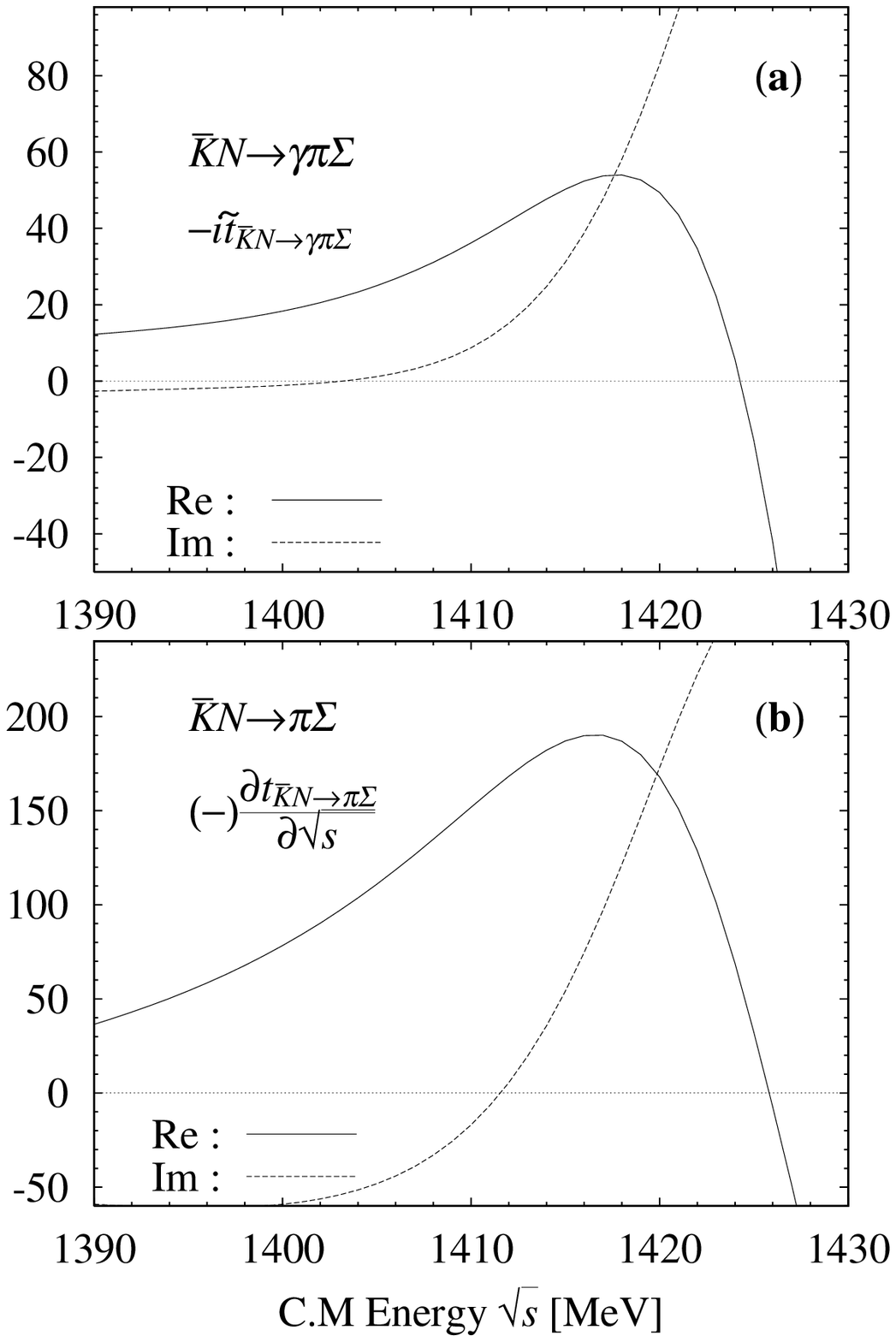}
   \end{center}
   \caption{ Real and imaginary parts of a) the numerator $-i \tilde
    t_{\bar K N \rightarrow \gamma \pi \Sigma}$ in eq.(\ref{realratio}) 
    and b) the denominator $-\del t_{\bar K N \rightarrow \pi \Sigma} /
    \del \sqrt{s}$ in eq.(\ref{realratio}) around the $\Lambda(1405)$
    resonance region in units of $m_\pi^{-2}$. \label{fig:plot1_2}} 
 \end{figure}
\begin{figure}
   \begin{center}
    \epsfxsize=10cm
    \epsfbox{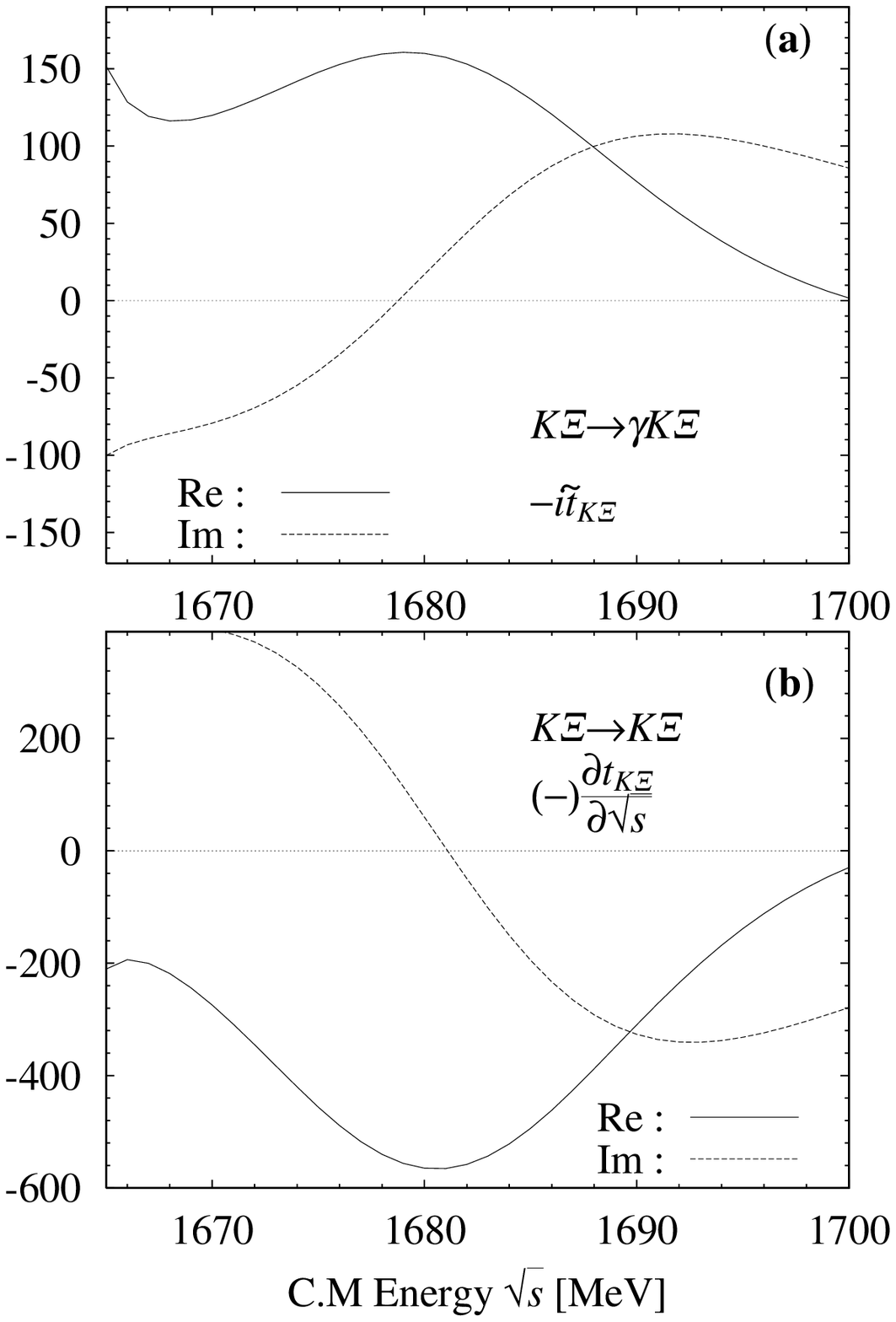}
   \end{center}
    \caption{Real and imaginary parts of a) the numerator 
    $-i \tilde t_{K \Xi}$ in eq.(\ref{realratio}) and b) the
   denominator $-\del t_{K \Xi} / \del \sqrt{s}$ in
   eq.(\ref{realratio}) around the $\Lambda(1670)$ resonance region in
   units of $m_\pi^{-2}$. \label{fig:plot2}} 
 \end{figure}

The analysis in the complex plane has the advantage of making the 
background contributions  negligible since the evaluations are
done exactly at the poles of the resonances. The magnetic moment
evaluated in the complex plane, however, has a complex value, which 
might induce uncertainties since one is extrapolating from the real
axis to the complex plane.
Hence,
to avoid these uncertainties, 
we also calculate the amplitudes on the real axis in the first Riemann 
sheet.  
The magnetic moments are then defined by 
\begin{equation}
    \mu_{\Lambda^{*}} = {- i \tilde t_{ij}(\sqrt{s}) \over - {\del
    \over \del \sqrt{s} } t_{ij}(\sqrt{s})}  \label{realratio} \ ,
\end{equation}
where both the coupling constants and the resonance propagators cancel 
to provide the magnetic moment of the resonance. In order to 
eliminate background we choose external channels which have a large 
coupling to the resonances and, furthermore, we take 
the $I=0$ isospin combination.  In particular, we take the $\bar
K N$ state with $I=0$ for $\Lambda(1405)$ and the $K\Xi$ state with $I=0$ for
$\Lambda (1670)$ because of their large couplings to the
corresponding  channels~\cite{cornelius}. 
For $\Lambda(1405)$ we
also calculate the magnetic moment in the $\bar K N
\rightarrow \gamma \pi \Sigma$ channel, since this channel may be used
in the experiments to determine the magnetic moment of the 
$\Lambda (1405)$. We show the 
numerator and the denominator of eq.(\ref{realratio})  in 
fig.\ref{fig:plot1} with the $\bar K N$ channel, in
fig.\ref{fig:plot1_2} with the $\bar KN \rightarrow \gamma\pi \Sigma$
for the $\Lambda(1405)$  and in fig.\ref{fig:plot2} with the $K \Xi$
channel for the $\Lambda(1670)$.  
We take the ratio of these amplitudes around
the energy close to the resonance where the real part of the two
functions has maximum strength. In order to estimate uncertainties we
also evaluate the ratio at the point where either the imaginary part of
the numerator or denominator becomes zero, as well as the ratio of the 
dominant real parts. In principle, in the absence of  background
contamination, these evaluations should give the same value.

\begin{figure}
   \begin{center}
    \epsfxsize=15cm
    \epsfbox{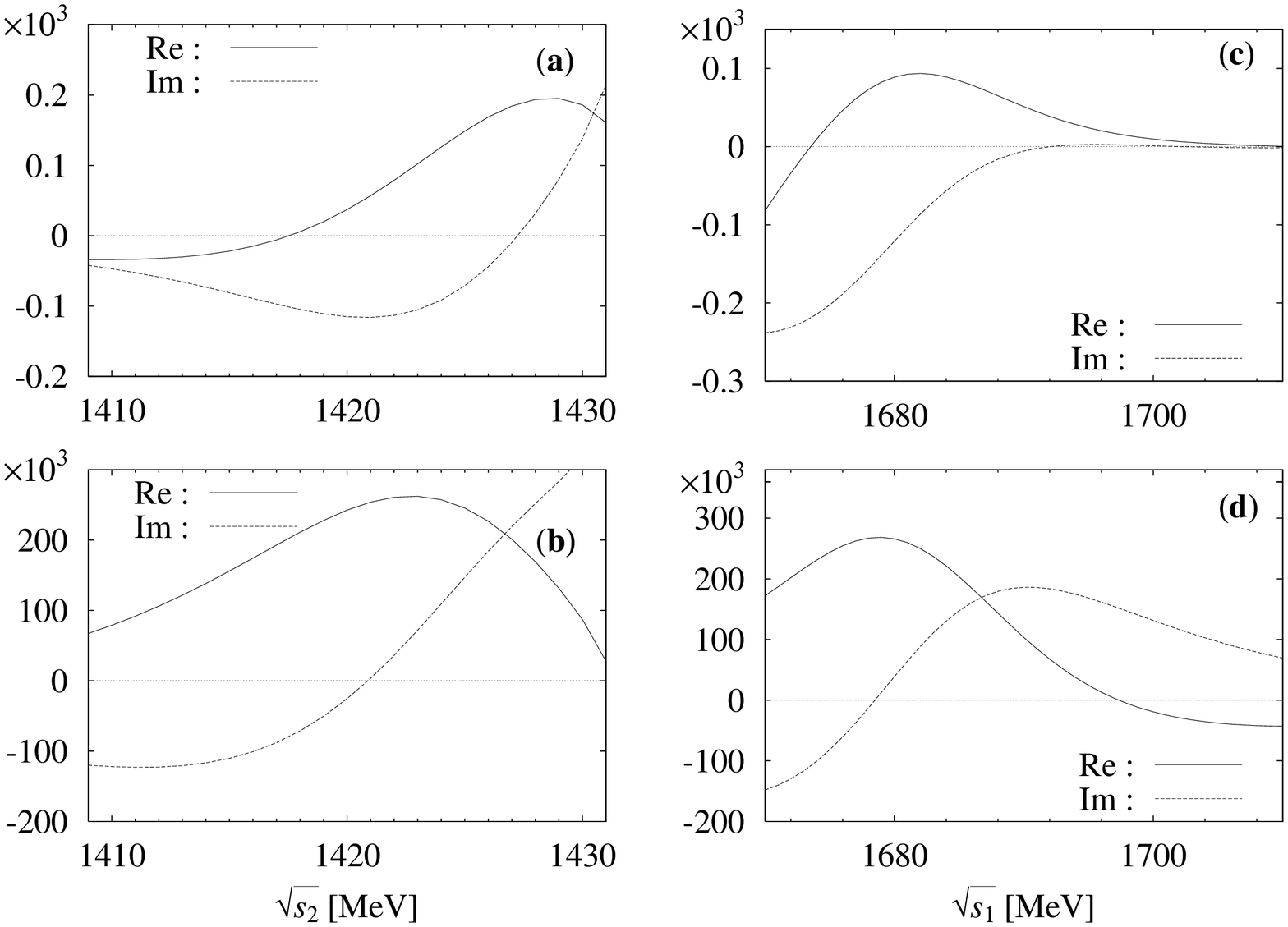}
   \end{center}
    \caption{Real and imaginary parts of the numerator, a) and c), and
    the denominator, b) and d), in eq.(\ref{tranmag}) in units of
    $m_\pi^{-4}$. In a) and b)
    $\sqrt{s_1}$ is fixed at 1680 MeV and the numerator and the
    denominator are functions of $\sqrt{s_2}$. In c) and d)
    $\sqrt{s_2}$ is fixed at 1420 MeV and the numerator and the
    denominator are functions of $\sqrt{s_1}$. \label{fig:plottrans}}
 \end{figure}

As for the transition magnetic moment, in order to 
cancel the couplings and propagators, we take the ratio
\begin{equation}
    \mu^{2}_{\Lambda(1670)\rightarrow\Lambda(1405)} = { (-i \tilde 
    t_{K\Xi\rightarrow \gamma \bar KN}(\sqrt{s_{1}}, \sqrt{s_{2}}) )
    (i \tilde t_{\bar KN \rightarrow \gamma K\Xi}(\sqrt{s_{2}},
    \sqrt{s_{1}}))  \over \left(- {\del \over \del \sqrt s} t_{K\Xi}
    (\sqrt {s_{1}})\right) 
    \left(- {\del \over \del \sqrt s} t_{\bar KN} (\sqrt
    {s_{2}})\right)} \label{tranmag}
\end{equation}
and we proceed as before to evaluate the ratio and the uncertainties. 
We show in fig.\ref{fig:plottrans} the numerator and the denominator of 
eq.(\ref{tranmag}), for fixed $\sqrt{s_{2}} = 1681$ MeV as a 
function of $\sqrt{s_{1}}$ in the left panels, and for fixed
$\sqrt{s_{1}} = 1423$ MeV as a function of $\sqrt{s_2}$ in the right panels.
    
Experimentally, magnetic moments of resonances may be extracted 
from bremsstrahlung processes, which are carefully compared with 
theoretical models.  
On the other hand, 
the transition magnetic moment between $\Lambda(1670)$ and 
$\Lambda(1405)$ could be directly investigated from the decay 
$\Lambda(1670) \rightarrow \Lambda(1405) \gamma$. The width for this 
transition is given by 
\begin{equation}
    \Gamma = {1 \over \pi} {M_{\Lambda(1405)} \over M_{\Lambda(1670)} }
    q^{3} \left( {e \mu_{\Lambda(1670)\rightarrow\Lambda(1405)} 
     \over 2 \protonmass} \right)^{2} 
    \label{decay}
\end{equation}
with $q$ the photon momentum in the $\Lambda(1670)$ rest frame.

\section{Results}

\begin{table}
  \begin{tabular}{l|ccc}
  \hline
     & $\Lambda(1405)$ & $\Lambda(1650)$ & transition \\
  \hline
  real axis & $+ 0.44 \pm 0.06^{a)} $ & $-0.29\pm0.01$ & $0.023\pm 0.009$
  \\
  & $+0.26\pm0.07^{b)}$ & &  \\
  \hline
  complex plane & $0.41\pm 0.01^{c)}$ & $0.23$ & $0.019\pm0.002^{c)}$ \\
  (absolute value) & $0.30\pm0.01^{d)} $ & & $0.093\pm0.003^{d)} $\\
  \hline
  \end{tabular}
  \caption{Magnetic moments obtained by the chiral unitary approach in
  units of the nuclear magneton. The values without signs denote the
  modules. a) calculation in the $\bar KN \rightarrow \gamma \bar KN$
  channel. b) calculation in the $\bar KN \rightarrow \gamma
  \pi\Sigma$ channel. c) taking $z_R = 1426 + 16i$ for
  $\Lambda(1405)$. d) taking $z_R = 1390 + 66i$ for
  $\Lambda(1405)$. \label{results}} 
\end{table}

Comparison of the numerator and denominator in eq.(\ref{realratio}) 
for the $\Lambda(1405)$ with the $\bar K N \rightarrow \gamma \bar K
N$ and $\bar KN \rightarrow \gamma \pi \Sigma$ channel 
and performing the ratios discussed in the 
former section we obtain a value
\begin{equation}
   \mu_{\Lambda(1405)} = +0.24 \sim 0.45  \label{result1} 
\end{equation}
in units of the nuclear magneton $\mu_N = e/2M_p$. 
The large uncertainty in the result obtained comes from the energy
range where the amplitudes of the ratio of eq.(\ref{realratio})
are evaluated. As seen in figs. \ref{fig:plot1} and \ref{fig:plot1_2}
the value of this energy, which signals the position of the resonance
in the real axis, is around 1418-1422 MeV for the $\bar KN$ channel
and 1403-1416 MeV for the $\pi\Sigma$ channel.
The evaluation in the $\bar K N$ channel gives
$\mu_{\Lambda(1405)} = +0.44 \pm 0.06$, while in the $\bar KN
\rightarrow \gamma \pi \Sigma$ we obtain $+0.26\pm 0.07$. 
We also evaluate the magnetic moment using the ratio of
eq.(\ref{compratio}) at the pole in the second Riemann sheet, which
gives a complex number with the module $0.41\pm 0.01$ for the case of
$z_R = 1426+16i$ and $0.30\pm 0.01$ for $z_R=1390+66i$. All possible
isospin $I=0$ combinations, $\bar K N$, $\pi \Sigma$, $\eta \Lambda$
and $K\Xi$, provide approximately the same value (the channel
dependence is shown in the small error bar of the presented value.) This
channel insensitivity in the evaluation in the complex plane implies
that the ratio of eq.(\ref{compratio}) at the pole is dominated by the
resonance and is not affected by background contaminations. It is
interesting to note that the values in the complex plane are comparable
with the value of eq.(\ref{result1}). In addition, recalling that the
pole at $z_R = 1426+16i$ couples largely to $\bar KN$ and that at
$z_R=1390+66i$ to $\pi \Sigma$, the channel (or energy) dependence of
the magnetic moment evaluated on the real axis
stems from a different contribution of each pole 
to the values of the amplitudes in the real axis.

For the case of the $\Lambda(1670)$ the ratio obtained from 
fig.\ref{fig:plot2} with the $K\Xi$ channel gives us 
\begin{equation}
    \mu_{\Lambda(1670)} = -0.29 \pm 0.01\label{result2}
\end{equation}
with small uncertainty, and we find that the ratio of eq. (\ref{realratio})
is stable around the resonance region.
It is also interesting to note that the analysis in the complex plane 
in the pole in the second Riemann sheet (eq.(\ref{compratio})) gives 
in this case a value for the modulus of $0.23$, which is similar to
that of eq.(\ref{result2}). As in the preceding case, 
the analysis in the real plane
allows us to obtain a real magnetic moment with a given sign.

Finally for the case of the transition magnetic moment we obtain the
value from eq.(\ref{tranmag}) and fig.\ref{fig:plottrans} 
\begin{equation}
    \left| \mu_{\Lambda(1670)\rightarrow \Lambda(1405)} \right| = 
    0.023\pm 0.009 \label{result3}
\end{equation}
We also evaluate the transition magnetic moment from
eq.(\ref{tranratio}) in the complex plane, which gives the modulus
$0.019\pm 0.002$ with $z_{1R} = 1680+20i$, $z_{2R} = 1426+16i$ and
$0.093\pm 0.003$ with $z_{1R} = 1680+20i$, $z_{2R} = 1390+66i$.
The values obtained in the complex plane 
are less reliable in this case because they involve an extrapolation 
of two variables to 
the complex plane, each of which induces
uncertainties. Even then, the agreement with the evaluation on the
real axis is fair if we take into account the fact that, given the
smallness of these numbers, their differences are of the same order of
magnitude than those for the $\Lambda(1405)$ case.
The results discussed here are summarized in table \ref{results}.

With the value of the transition magnetic moment of 
eq.(\ref{result3}) and using eq.(\ref{decay}) we obtain a partial
width for the $\Lambda(1670) \rightarrow \Lambda(1405)\gamma$ decay 
which corresponds to a branching ratio $2 \times 10^{-6}$.

\section{Quark Model Results}

In this section we compute the resonance magnetic moments in the
non-relativistic quark model.
This demonstrates that the nature of the resonances differ
appreciably from the chiral unitary description.
In the $SU(6)$ quark model, 
the $\Lambda(1405)$ and $\Lambda(1670)$ are described as 
$p$-wave excitations of the 70-dimensional
representation, whose $SU(2) \times SU(3)$ decomposition is given by
\beq
70 = {^2 8} + {^4 8} + {^2 1} + {^2 10} \, .
\label{decomp70}
\eeq
Here in the notation on the right hand side,
$^{2j+1} D$,  $j$ represents the resonance spin and $D$
the dimension of 
the flavor $SU(3)$ representation.

Since the $\Lambda$ particles are isosinglet, their wave functions
are spanned by the flavor octet and singlet states.
Explicitly, these states are given as~\cite{ahbook}
\beq
|^{2}8; jm\ket &=& 
\half 
\left( 
[\psi(\rho), \chi_{\rho}]_{jm} \phi_{\lambda}
+ [\psi(\rho), \chi_{\lambda}]_{jm} \phi_{\rho} \right. \nonumber \\
&&  \left. + [\psi(\lambda), \chi_{\rho}]_{jm} \phi_{\rho}
+ [\psi(\lambda), \chi_{\lambda}]_{jm}  \phi_{\lambda}
\right) \, , \nonumber \\
|^{4}8; jm\ket &=& 
\frac{1}{\sqrt{2}} 
\left( 
[\psi(\lambda), \chi_{S}]_{jm} \phi_{\lambda}
+ [\psi(\rho), \chi_{S} ]_{jm} \phi_{\rho}
\right)  \, , 
\label{qmwf}\\
|^{2}1; jm\ket  &=&
\frac{1}{\sqrt{2}} 
\left( 
[\psi(\lambda), \chi_{\rho}]_{jm}
- [ \psi(\rho) , \chi_{\lambda}]_{jm}
\right) \phi_{A} \nonumber
\, .
\eeq
Here we have employed standard notations:
\beq
& & \vec \rho = \frac{1}{\sqrt{2}}(\vec x_{2} - \vec x_{1})
\nonumber \\
& & \vec \lambda = \frac{1}{\sqrt{6}}
(\vec x_{2} + \vec x_{1} -2 \vec x_{3})
\nonumber \\
& & \psi(\vec x) : \; \;
p \; {\rm wave \;\; orbital \;\; wave \;\; functions}
\nonumber \\
& & \chi_{\rho, \lambda, S} : \; \;
{\rm flavor \;\; wave \;\; functions \;\; of}\;\; \rho, \;\;
\lambda \;\; {\rm and} \;\; S \;\; {\rm symmetry}
\nonumber \\
& & \phi_{\rho, \lambda, A} : \; \;
{\rm flavor \;\; wave \;\; functions \;\; of}\;\; \rho, \;\;
\lambda \;\; 
{\rm and} \;\; A \;\; {\rm symmetry}
\nonumber
\eeq
Furthermore, 
in eq. (\ref{qmwf}), the orbital and spin wave functions are coupled to the
total spin $jm$.  

In the non-relativistic description, the magnetic moment operator is
given by the sum of twice the spin and the orbital angular momentum:
\beq
\vec \mu = \frac{e}{2m}\sum_{i=1}^{3}
\left(
\vec \sigma (i) + \vec l (i)
\right)
\left(
\half \lambda_{3}(i) + \frac{1}{2\sqrt{3}}  \lambda_{8}(i)\right) \, .
\eeq
In this equation $m$ is a constituent quark mass for which we take for 
simplicity a 
common value $m \sim M_{N}/3$ for the three quarks.  
Furthermore, we have written the charge matrix as a sum of $SU(3)$ 
components.  
Due to the isosinglet nature of the $\Lambda$ particles, the matrix
elements of the isovector 
($\lambda_{3}$) term vanish and only the $\lambda_{8}$-term
contributes.  
The actual computation is straightforward and therefore here we present only
the final result.  
By writing a $\Lambda$ state as
\beq
|\Lambda\ket =
a_{1} | ^2 8\ket + a_{2} | ^4 8\ket + a_{3} | ^2 1 \ket \, ,
\label{lmdexp}
\eeq
where the coefficients must satisfy the normalization condition,
$a_{1}^2 + a_{2}^2 + a_{3}^3 = 1$, we find for the diagonal
element:
\beq
\bra \Lambda |\mu_{z}|\Lambda \ket
&=& - \frac{1}{12} a_{2}^2  + \frac{1}{9} a_{1}a_{3} \, .  
\eeq
Similarly we can compute the off-diagonal matrix element.  

The coefficients are determined by assuming suitable
interactions between quarks.  
Here we employ two parameter sets; Isgur-Karl (IK)~\cite{isgurkarl}
and Hey-Litchfield-Cashmore (HLC)~\cite{hey}, the values of which are
shown in Table~\ref{coeff}.
We summarize the results for the magnetic moments
in Table~\ref{qmresults}, where the results for the transition
magnetic moment are also shown.  
We find that the magnetic moment of the $\Lambda(1405)$ is  small 
and negative, as opposed to the result of the chiral unitary 
method.  
For the $\Lambda(1670)$ the HLC parameters provide a value similar to the 
chiral unitary method.  
For the transition magnetic moment, however, the quark model values 
are significantly larger than the chiral unitary ones, by about a 
factor ten.  
This in turn results in the branching ratio in the $\Lambda(1670)$ 
decay about hundred times larger than the chiral unitary result.  
In general, the magnetic moments are sensitive to the choice of 
the mixing coefficients.  
However, absolute values are small for a reasonable range of the 
mixing parameters.
In the quark model, there are two reasons for this:
\begin{enumerate}
    \item Only the isoscalar ($\lambda_{8}$) 
    component contributes, which has relatively
    small contributions.
    \item For spin doublet $^2 8$ and $^2 1$ states, the spin $\vec s$ and
    orbital angular momentum $\vec l$ are aligned such that their
    contributions roughly cancel.
    Classically, to form a $(jm) = (1/2, 1/2)$ state,
    $\vec l$ orients to the $+z$ direction, while $\vec s$ orients to
    the $-z$ directions;
    $\bra l_{z}\ket  = 1$ and $\bra s_{z}\ket = -1/2$.
    The magnetic moment of this configuration vanishes:
    \beq
    \bra \mu_{z} \ket = 1 \bra l_{z} \ket + 2 \bra s_{z} \ket = 0 \, .
    \eeq
\end{enumerate}

\begin{table}[tbp]
  \centering
  \begin{tabular}{c c c c | c c c}
   \hline
                &   & $\Lambda(1405)$  & & &  $\Lambda(1670)$ &  \\
  & $a_{1}$ & $a_{2}$ & $a_{3}$ & $a_{1}$ & $a_{2}$ & $a_{3}$ \\
   \hline
  IK & 0.43 & 0.06 & 0.9  & 0.75 & 0.58 & $-0.39$ \\
 HLC & 0.46 & 0.25 & 0.85 & $-0.04$ & $-0.95$ & 0.30  \\
   \hline
  \end{tabular}
  \caption{Expansion coefficients in (\ref{lmdexp}). \label{coeff}}
 \end{table}

 \begin{table}[tbp]
  \centering
  \begin{tabular}{c c c c}
   \hline
     & $\Lambda(1405)$  & $\Lambda(1670)$ &
               $|\Lambda(1405)$-$\Lambda(1670)|$\\
   \hline
  IK & $-0.13$ & 0.01 & 0.14   \\
  HLC & $-0.15$ & $-0.23$ & 0.26   \\
   \hline
  \end{tabular}
  \caption{Magnetic moments in the quark model in units of nuclear
  magneton $e/2m_{p}$. \label{qmresults}}
 \end{table}

\section{Conclusion}

We have introduced here the formalism to evaluate magnetic moments
and the transition magnetic moment of the two $\Lambda^{*}$ resonances,
$\Lambda(1405)$ and $\Lambda(1670)$, which are dynamically generated
within $U\chi PT$.  At the same time we have done the numerical 
evaluations and have determined the actual value for these magnitudes. The 
values obtained are $\mu_{\Lambda(1405)} = +0.2 \sim 0.5 \mu_{N}$, 
smaller than 
that of the $\Lambda$ ($\sim - 0.6 \mu_{N}$) and of opposite sign. For 
the $\Lambda(1670)$ we obtain $\mu_{\Lambda(1670)} \sim -0.29 
\mu_{N}$, also smaller than that of the $\Lambda$ and with the same sign, 
while for the transition magnetic moment we obtain a value $| 
\mu_{\Lambda(1670)\rightarrow \Lambda(1405)} | \sim  0.023 \mu_N$, 
which leads to a branching ratio of the $\Lambda(1670)$ to 
$\Lambda(1405) \gamma$ channel of the order of $2\times 10^{-6}$.  The 
results of the $U\chi PT$ method are different from those 
obtained with the quark models, reflecting the different nature
attributed to the resonances in those models.
One of the interesting results obtained in
this work is the abnormally small decay width for the $\Lambda(1670)
\rightarrow \Lambda(1405) \gamma$ transition, which differs in two
orders of magnitude from the quark model predictions. Short of a
measurement of the transition, which could be difficult 
given the small numbers
predicted, even the determination of an upper bound would provide
interesting information about the nature of these resonances.

\subsection*{Acknowledgments}

E. O. and D.J. wish to acknowledge the hospitality of the University
of Barcelona. 
A. H.  thanks the members of the 
nuclear theory group at the University of Valencia, where part 
of this work was done.  This work is also
partly supported by DGICYT contract numbers BFM2000-1326, PB98-1247,
by the EU TMR network Eurodaphne, contract no. ERBFMRX-CT98-0169, and
by the Generalitat de Catalunya project SGR2000-24.

\newpage

\begin{thebibliography}{99}

\bibitem{isgurkarl} 
     N. Isgur and G. Karl, Phys. Rev. D18 (1978) 4187; 
     {\it ibid.} D20 (1979) 1191; 
     S. Capstick and W. Roberts, Phys. Rev. D49 (1994) 4570.
%
\bibitem{pedro} 
     F. Cano, P. Gonzalez, S. Noguera, B. Desplanques, 
     Nucl. Phys. A603 (1996) 257.
%
\bibitem{Kai95} 
     N. Kaiser, P. B. Siegel and W. Weise, 
     Nucl. Phys. A594 (1995) 325.
%
\bibitem{angels} 
     E. Oset and A. Ramos, Nucl. Phys. A635 (1998) 99.
%
\bibitem{joseulf} 
     J.A. Oller and U.G. Meissner, 
     Phys. Lett. B500 (2001) 263.
%
\bibitem{Kai97} 
     N. Kaiser, T. Waas and W. Weise, Nucl. Phys. A612 (1997) 297.
%
\bibitem{Nacher:2000vg}
     J.~C.~Nacher, A.~Parreno, E.~Oset, A.~Ramos, A.~Hosaka and M.~Oka,
     Nucl.\ Phys.\ A {\bf 678} (2000) 187.
%
\bibitem{Nieves:2001wt}
     J.~Nieves and E.~Ruiz Arriola,
     Phys.\ Rev.\ D {\bf 64} (2001) 116008.
%
\bibitem{Inoue:2001ip}
     T.~Inoue, E.~Oset and M.~J.~Vicente Vacas,
     Phys.\ Rev.\ C {\bf 65} (2002) 035204.
%
%
\bibitem{cornelius} 
     E. Oset,  A. Ramos and C. Bennhold, 
     Phys. Lett. B527 (2002) 99-105.
%
\bibitem{angelsnow} A. Ramos, E. Oset and C. Bennhold, 
     in preparation.
%
\bibitem{Oller:1997ti}
     J.~A.~Oller and E.~Oset,
     Nucl.\ Phys.\ A {\bf 620} (1997) 438
     [Erratum-ibid.\ A {\bf 652} (1997) 407].
%
\bibitem{daniel} 
     D. Cabrera, E. Oset, M.J. Vicente Vacas, Nucl. Phys. A, in
     print, nucl-th/0011037.
%
\bibitem{Be95} 
     V. Bernard, N. Kaiser and U. G. Meissner, 
     Int. J. Mod. Phys. E4 (1995) 193.
%
\bibitem{Pi95} 
     A. Pich, Rep. Prog. Phys. 58 (1995) 563.
%
\bibitem{Meissner:1997hn}
     U.~G.~Meissner and S.~Steininger,
     Nucl.\ Phys.\ B {\bf 499} (1997) 349.
%
\bibitem{gl} 
     J. Gasser and H. Leutwyler, Nucl. Phys. B250 (1985) 465.
%
\bibitem{Nacher:1999ni}
     J.~C.~Nacher, E.~Oset, H.~Toki and A.~Ramos,
     Phys.\ Lett.\ B {\bf 461} (1999) 299.
%
\bibitem{hey}  
    A.J.G. Hey, P.J. Litchfield and R.J. Cashmore, 
    Nucl. Phys. B95 (1975) 516.
%
\bibitem{ahbook}
    See for example, A. Hosaka and H. Toki,
    ``Quarks, baryons and chiral symmetry'', World Scientific (2001).

\end{thebibliography}
\end{document}